\documentclass[%
 reprint,
 amsmath,amssymb,
 aps,
]{revtex4-2}

\usepackage{graphicx}

\begin{document}


\title{Ionization-induced Susceptibility by Nearly-free Electrons in Gases Influenced by the Coulomb Potential}

\author{F. R. Arteaga-Sierra}
\author{J. Herrmann}
\author{A. Husakou}
 \email{gusakov@mbi-berlin.de}
\affiliation{
 Max Born Institute, Max Born Str. 2a, 12489 Berlin, Germany}%


\begin{abstract} In the present paper we study the influence of the Coulomb potential on the real and imaginary parts of the plasma-induced susceptibility in a photoionized gas. We show that the real part of the susceptibility is more than one order of magnitude larger due to the action of a Coulomb potential. Surprisingly, the long-range Coulomb potential of the atomic core leads to an additional contribution to the imaginary part of the susceptibility which has no counterpart in the case of a short-range potential. We demonstrate that the origin of this behavior are electrons in states very close to the continuum (nearly-free electrons), and analyze the dependence of the susceptibility on the intensity and wavelengths.

\end{abstract}

\maketitle


\section{\label{sec:level1}Introduction}

In recent years, the development of new mid-infrared (3–8 $\mu$m) and longwave-IR (8-15 $\mu$m) laser sources has received significant attention due to numerous possibilities for both fundamental  and applied research \cite{Pires, Wolter}. Mid-IR femtosecond pulses with high energies can be efficiently generated by optical parametric chirped pulse amplification \cite{Rothhard} using different schemes via parametric down conversion \cite{Hemmer, Hong, Liang, Hong2, Grafenstein, Fuertjes}.

The mid-infrared spectral range differs in several important characteristics from the near-IR or visible spectral range.  The dominant physical effects in the interaction of ultrashort pulses with matter are dispersion and various contributions to the nonlinearity, originating mainly from the Kerr effect and from the plasma contribution of free ionized electrons in the gas. While the Kerr part is only weakly wavelength-dependent, the negative contribution of the plasma quadratically increases with the wavelength. Electrons driven by intense ultrashort mid-IR field waveforms acquire unusually high ponderomotive energies within a fraction of the field cycle. The self-focusing threshold increases with the square of the wavelength, allowing much higher peak powers to be transmitted in a single mode in the mid-IR range without loosing spatial coherence. Recently, unusual self-guiding of long-wave infrared CO$_2$ picosecond terawatt laser pulses with 2.5 J energy (megafilaments) has been observed \cite{Tochitsky} over a distance of 30 meters. One explanation of this phenomenon is that self-focusing collapse is arrested by many-body Coulomb-induced ionization \cite{Schuh,Tochitsky}. However, Ref. \cite{Woodbury} casts doubt on this explanation and explains the megafilaments by avalanche ionization seeded by the action of low-concentration  aerosols in air. 
The peculiarities of the mid- and longwave-infrared range require further theoretical understanding of the complex interactions of high-intensity ultrashort mid-IR pulses with matter. 

One of the main theoretical approaches for the description of atomic and molecular processes in intense laser fields is the so-called strong-field approximation (SFA)\cite{Keldysh, Faisal, Reiss}.  In the original form of this approach the effect of Coulomb interaction of the ionized electrons is neglected.  However, the obtained results were in striking contradiction with mid-infrared experiments, which show spike-like low-energy structures in the energy distribution of the ionized electrons.  Nonperturbative influence of the Coulomb effect on the electron dynamics \cite{blaga} has to be taken into account to explain this feature, which can be achieved by various extensions of the SFA. One of these approaches is the Coulomb-corrected SFA \cite{Popruzhenko} describing the electron continuum state in the SFA by the eikonal  Coulomb-Volkov state. An alternative approach is the analytic R-matrix method (ARM) \cite{Torlina}  in which the space is separated into an inner and an outer region and the solution is matched at the boundary. The latter approach is suitable for a numerically efficient calculation of the time-dependent polarization. 

The physical phenomena related to the electrons moving relatively far from the ionic core but still under the action of the long-range Coulomb potential has received significant attention in recent years. The dynamics of these nearly-free electrons is related to the Kramers-Henneberger states \cite{kulander,kh}, and contributes to various strong-field effects in gases, including stabilization of atoms \cite{eichmann2,eichmann5}, important contributions to higher-order nonlinearities \cite{morales}, as well as inversion and lasing \cite{nature_ivanov}. However, the role of the nearly-free electrons and peculiarities of their dynamics in the Coulomb potential, in particular with respect to the plasma-induced contribution to the refractive index, was not studied up to now.

In the present paper we study the contribution of ionization to the nonlinear refraction in gases in the midinfrared region taking into account the long-range Coulomb interaction by the ARM method. The paper is organized as follows: in Section II, the theoretical model is presented. In Section III, we start with discussion of the frequency-dependent plasma contribution to dispersion, followed by the detailed analysis of the nearly-free electrons in the Coulomb potential and, for comparison, a short-range potential. Finally, the conclusions of the current investigation are provided in Section IV.

\section{\label{sec:level2}Model for the description of the polarization}
In our theoretical approach we apply the method of analytical R-matrix (RAM) \cite{Torlina} which extends the SFA by including the Coulomb interaction of the ionized electron and the ionic core. First, for the sake of self-contained representation, we give the key formalism of this approach, closely following Ref. \cite{Torlina}. The main idea of this method consists of partitioning the space into an outer (sufficiently far from ion) and an inner (close to the ion) zone. In both zones, significant simplifications become possible: while in the outer zone the electron motion can be considered as free, the limited volume of the inner zone allows for analytical solutions of the bound states. This method reflects the different dynamics of the two regions into a compact and numerically tractable formula for the ionization amplitude. Although such approach neglects  few effects (among others, multiple hard recollisions), it was shown to be accurate for a broad range of parameters \cite{Torlina}.

We use the time-dependent ionization amplitude $a_{\mathbf{p}}(t)$ for the photo-electron emission with canonical momentum of $\mathbf{p}$  
      	\begin{equation}
      	\begin{aligned} a_{\mathbf{p}}(t)=\sum_{t_s}&  R_{\kappa l m}(\mathbf{p}) e^{-i \int_{t_{k}}^{t} d \tau U\left(\int_{t s}^{\tau} d t^{\prime \prime}\left[\mathbf{p}+\mathbf{A}\left(t^{\prime \prime}\right)\right]\right)} \\ & \times e^{-\frac{i}{2} \int_{t_s}^{t} d \tau[\mathbf{p}+\mathbf{A}(\tau)]^{2}+i I_{p} t_{s}.} \end{aligned}
      	\label{Eq:IonAmp}
      	\end{equation} \\
Here, $k,l,m$ are the indices of the representation of the ground-state wavefunction in terms of hydrogenic states, $U(\mathbf{r})$ is the Coulomb potential, $I_p$ is the ionization potential, $t_k=t_s-i/2I_p$, $t_s$ are the complex-valued ionization times, and $\mathbf{A}(t)$ is the vector potential of the field fulfilling $F(t)=-{d\mathbf{A}}/{dt}$. The pre-factor $R_{\kappa l m}$ is defined as	
      	\begin{equation}
R_{\kappa l m}(\mathbf{p})=\frac{(-1)^{l+m} i^{m+1}}{\sqrt{\left|S_{V}^{\prime \prime}\left(t, t_{s}\right)\right|}} C_{l m} e^{i m \phi_{p}}\left[\frac{p_{\perp}}{\kappa}\right]^{m} C_{\kappa l} \kappa^{1 / 2}
       \label{Eq:Rklm}
      	\end{equation}\\
with $\kappa=\sqrt{2I_p}$ and\\
      	\begin{equation}
C_{l m}=\frac{1}{2^{m} m !} \sqrt{\frac{(2 l+1)(l+m) !}{4 \pi(l-m) !}}.
       \label{Eq:Clm}
      	\end{equation}\\
       $C_{\kappa l}$ is the constant that fulfills the condition
       	\begin{equation}
\int|C_{\kappa l}\kappa^{3/2}\frac{e^{-\kappa r}}{\kappa r}(\kappa r)^{Q/\kappa}|^2dr=1.
       \label{Eq:Ckl}
      	\end{equation}\\     	
Eq. \ref{Eq:Rklm} includes the action of a particle moving with canonical momentum $S_{V}(\mathbf{p})$:
      	\begin{equation}
S_{V}\left(t, t^{\prime}, \mathbf{p}\right)=\frac{1}{2} \int_{t'}^{t}  v_{z}(\tau)^{2}d \tau-[I_{p}+p_\perp^2/2] t^{\prime}.
       \label{Eq:Action} 
      	\end{equation} \\ 
where $p_\perp$ is the component of the canonical momentum perpendicular to the direction of the electric field. 

The first exponential in Eq.  \ref{Eq:IonAmp} is the contribution of the Coulomb potential, and the second exponential takes into account the angular structure of the wave function. The values of $t_s$ are found using the saddle-point approximation, which provides  
\begin{equation}
      	\frac{1}{2}|\mathbf{p}+\mathbf{A}\left(t_s\right)|^{2}=-I_{\mathrm{P}}.       
      	\label{Eq:SaddleEq} 
      	\end{equation} \\ 
The gradient descent method is applied for the numerical solution of the above equation.
      	Saddle points are used to evaluate the integrals of the ionization amplitude formula (Eq.  \ref{Eq:IonAmp}). 
 The standard way for evaluating the integrals of Eq. \ref{Eq:IonAmp} is to start at the complex ionization time $t_s$, integrate directly to its real part $Re(t_s)$ on the real axis, and then along the real axis until the detection time $t$. 
 
The photoionization-induced current is then calculated by: 
      	\begin{equation}
J(t)=-e v(t)N_{L},     	
\label{Eq:Current} 
      	\end{equation}\\
 where $e$ is the electric charge, $N_{L}$ is the Loschmidt number, and\\
       	\begin{equation}
 v(t)=\sum_{ \mathbf{p}_\parallel, \mathbf{p}_\perp }  |a_{\mathbf{p}}(t)|^2[\mathbf{p}_\parallel-\mathbf{A}(t)] 2\pi |\mathbf{p}_\perp| d \mathbf{p}_\parallel d \mathbf{p}_\perp.
      	\label{Eq:Vel} 
      	\end{equation} \\
 Finally, the quantity which we are interested in is the susceptibility as a function of frequency, given in atomic units by
        \begin{equation}
 \chi(\omega)=4\pi \frac{P(\omega)}{F(\omega)},
  	\label{Eq:chi} 
      	\end{equation}
      	 where $F(\omega)$ is the Fourier transform of the input electric field $F(t)$. 
To avoid numerical artifacts due to non-zero values of the polarization at the end of the numerical domain, we obtain the Fourier transform of the polarization $P(t)=\int_{-\infty}^t J(t)dt$ from
      	\begin{equation}
            \mathcal{F}[\frac{d^2P(t)}{dt^2}]=-\omega^2P(\omega),   \label{Eq:FourierProp}
      	\end{equation}\\
where $\mathcal{F}$ denotes the Fourier transform. 

The above expression includes only the susceptibility due to photoionization process and due to the dynamics of the resultant plasma; the linear contribution to dispersion by bound electrons, nonlinear perturbative terms such as the optical Kerr effect, and terms describing the three-step harmonic generation need to be included additionally. Since in this paper we predominantly investigate the plasma-related terms, other above-mentioned terms are not written explicitly here.

\section{Results and discussion}

\begin{figure}[!ht]
\includegraphics[width=0.5\textwidth]{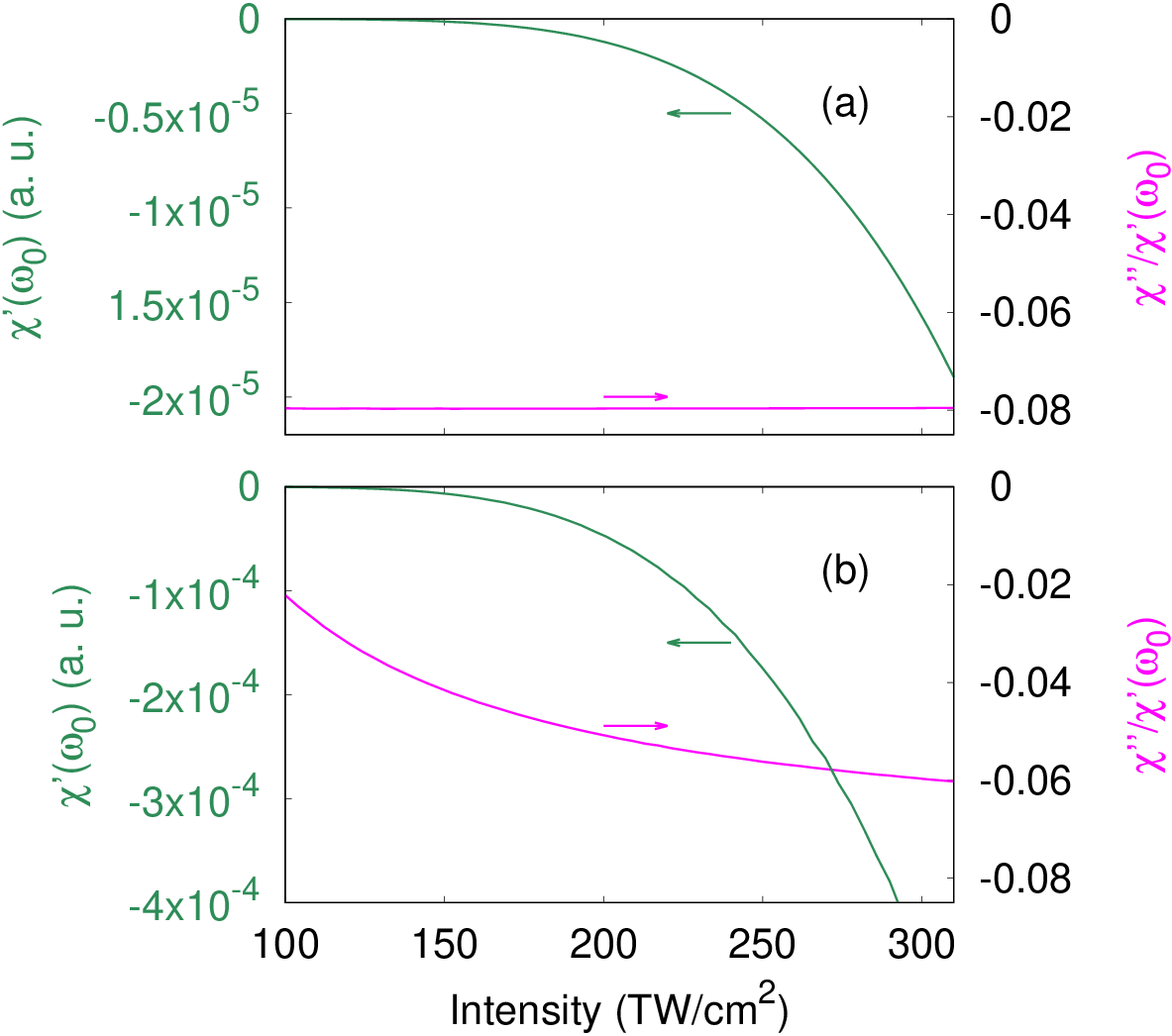}
\caption{\label{chi_int} Real (green curves, left axes) and imaginary parts (magenta curves, right axes) of the susceptibility at the pump frequency, for short-range (a) and long-range (b) potential. Atomic hydrogen with ionization potential of 0.5 a.u. at atmospheric pressure was considered.}
\end{figure}

We start the discussion by analyzing the intensity dependence of the ionization-related susceptibility of atomic hydrogen at the pump frequency corresponding to 2000 nm, taking into account the long-range Coulomb potential [Fig. 1(a)] as well as neglecting the Coulomb potential [Fig. 1(b)] for parameters indicated in thecaption. One can see that in both cases, the real part of the susceptibility (green curves) is negative, since it originates from the free-electron plasma, and strongly grows with intensity as expected due to the threshold-like behavior of the ionization rate and the plasma density. The real part of the susceptibility is more than one order of magnitude higher if the long-range Coulomb potential is included compared to short-range interaction without  the Coulomb potential. 

The real part of the susceptibility is associated with the change of the dielectric function due to the presence of the plasma. The imaginary part of the susceptibility is expected to be responsible for the losses, which can occur both due to scattering on atoms/ions as well as due to the direct energy absorption which happens during the photo-ionization process. Since the strength of all of these processes is proportional to the same quantity -- the ionization rate, or equivalently the plasma density, one might expect that the {\it ratio} $\chi'/\chi''$ between the real and imaginary parts of the susceptibility is intensity-independent, all other parameters being the same. 

Indeed, if we neglect the Coulomb potential, as shown in Fig. 1(a), the ratio $\chi'/\chi''$ remains constant. In contrast, one can see that this ratio, as shown by the magenta curve in Fig. 1(b), is intensity-dependent and lower if the Coulomb potential is included.  

\begin{figure}[!ht]
\includegraphics[width=0.5\textwidth]{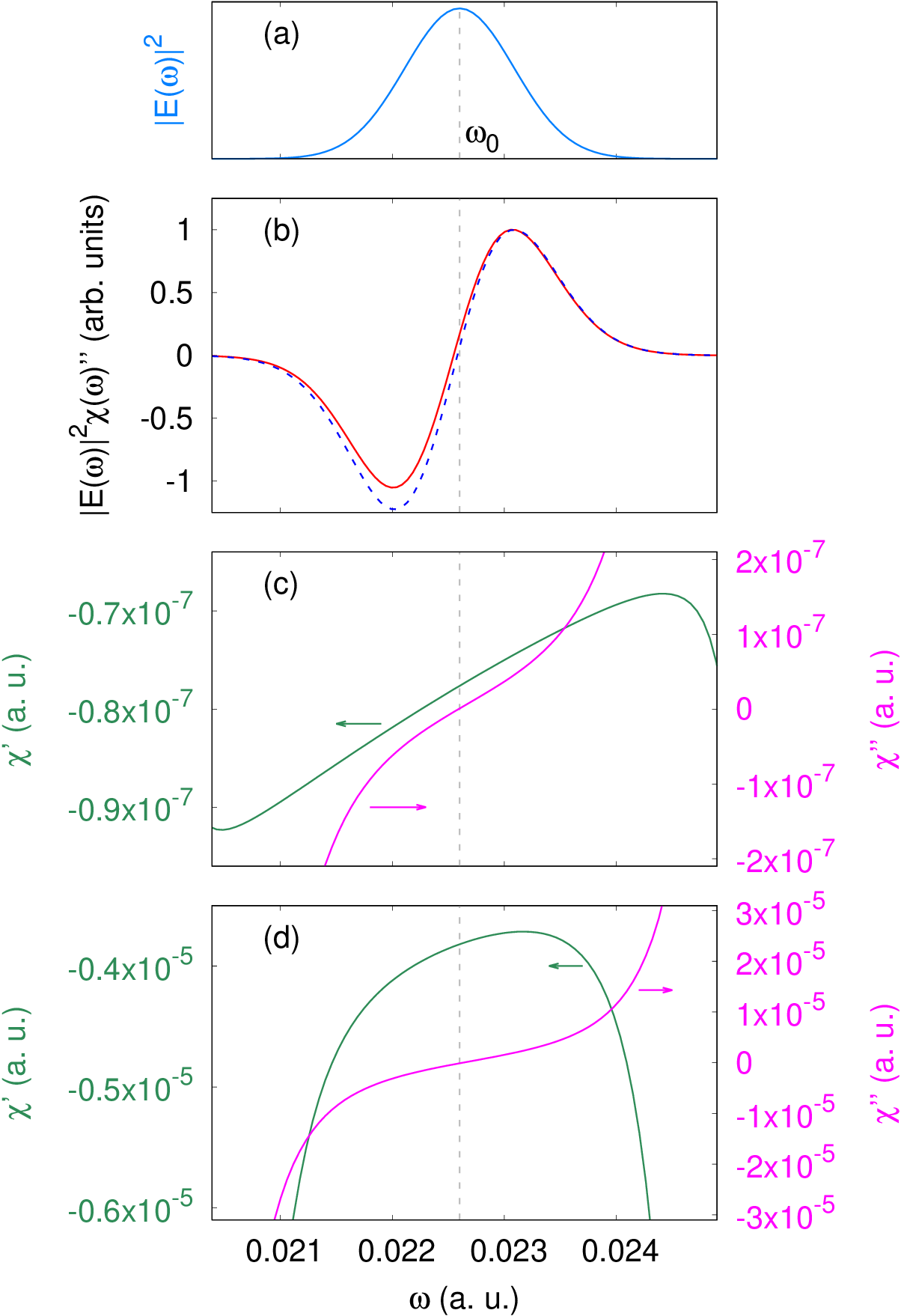}
\caption{\label{freq_dep} Frequency dependence of the susceptibility of atomic hydrogen at 1 atm for intensity of 100 TW/cm$^2$, without (c) and with (d) the Coulomb potential. In (b), the rates of the spectral intensity change $|E(\omega)|^2\chi''(\omega)$ are presented both with (red solid curve) and without the long-range Coulomb potential (blue dashed curve). In (a), the spectrum of the pump pulse is given.}
\end{figure}

However, a conclusion that due to Coulomb potential photo-ionization is accompanied by weaker nonlinear loss would be premature, as we up to now have considered the susceptibility only at the central frequency.
 In Fig. 2(c) and (d), we present the frequency dependence of the real and imaginary parts of the susceptibility both for the long-range Coulomb poential and the short-range potential. One can see that both the real and, in particular, the imaginary part of the susceptibility are strongly wavelength-dependent within the spectrum of the pump pulse [Fig. 2(a)]. Indeed, the imaginary part of the susceptibility changes sign in the vicinity of the pump frequency. 

This sign change implies that during propagation, the lower-frequency part of the pulse spectrum will be absorbed while the higher-frequency part will be amplified. This corresponds to the known effect of the spectral blue shift in the presence of a plasma. To further elucidate the spectral dynamics, we consider the frequency-dependent rate at which the spectral density changes with propagation. It is given by $\tilde{\alpha}(\omega)=|E(\omega)|^2\chi''(\omega)$ and shown in Fig. 2(b) for both without Coulomb effect (solid red curve) and with account of the Coulomb effect (dashed red curve) potentials. The curves were normalized so that their maximal value coincide on the plot. 

Without Coulomb effect the loss below $\omega_0$ and the gain above $\omega_0$ are almost balanced, and the integral of $\tilde{\alpha}(\omega)$ over the whole spectrum is close to zero. In fact this integral is of course negative, since the energy of the pulse is absorbed during the photoionization process to overcome the ionization potential $I_p$.

The above mentioned balance is, however, pronounced to a much lesser extent for the long-range Coulomb potential. There is a clear difference between the two curves in Fig. 2(b): for the frequencies below the pump frequency, the losses for the Coulomb potential are manifested stronger than for the short-range potential, and the balance is to a significant extent broken. As a result, with propagation a pulse will experience a much stronger absorption with the Coulomb effect than without it. 

\begin{figure}[!ht]
\includegraphics[width=0.5\textwidth]{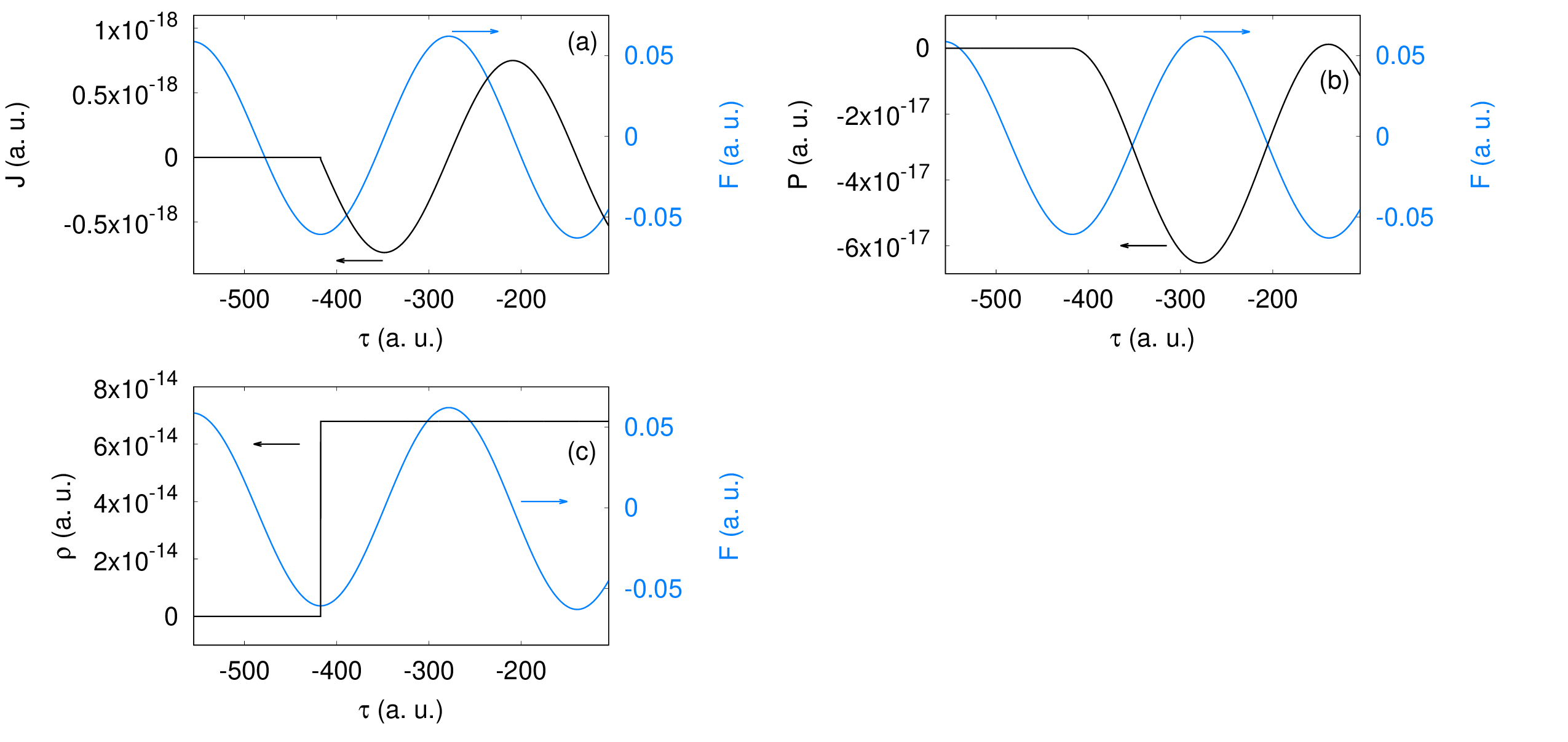}
\caption{\label{Iso0} Temporal dependencies of the plasma-induced current (a), polarization (b), and plasma density (c) for an isolated ionization event and the case without the Coulomb potential. In all cases, atomic hydrogen is considered, and blue curves represent the pump electric field.}
\end{figure}

\begin{figure}[!ht] 
\includegraphics[width=0.5\textwidth]{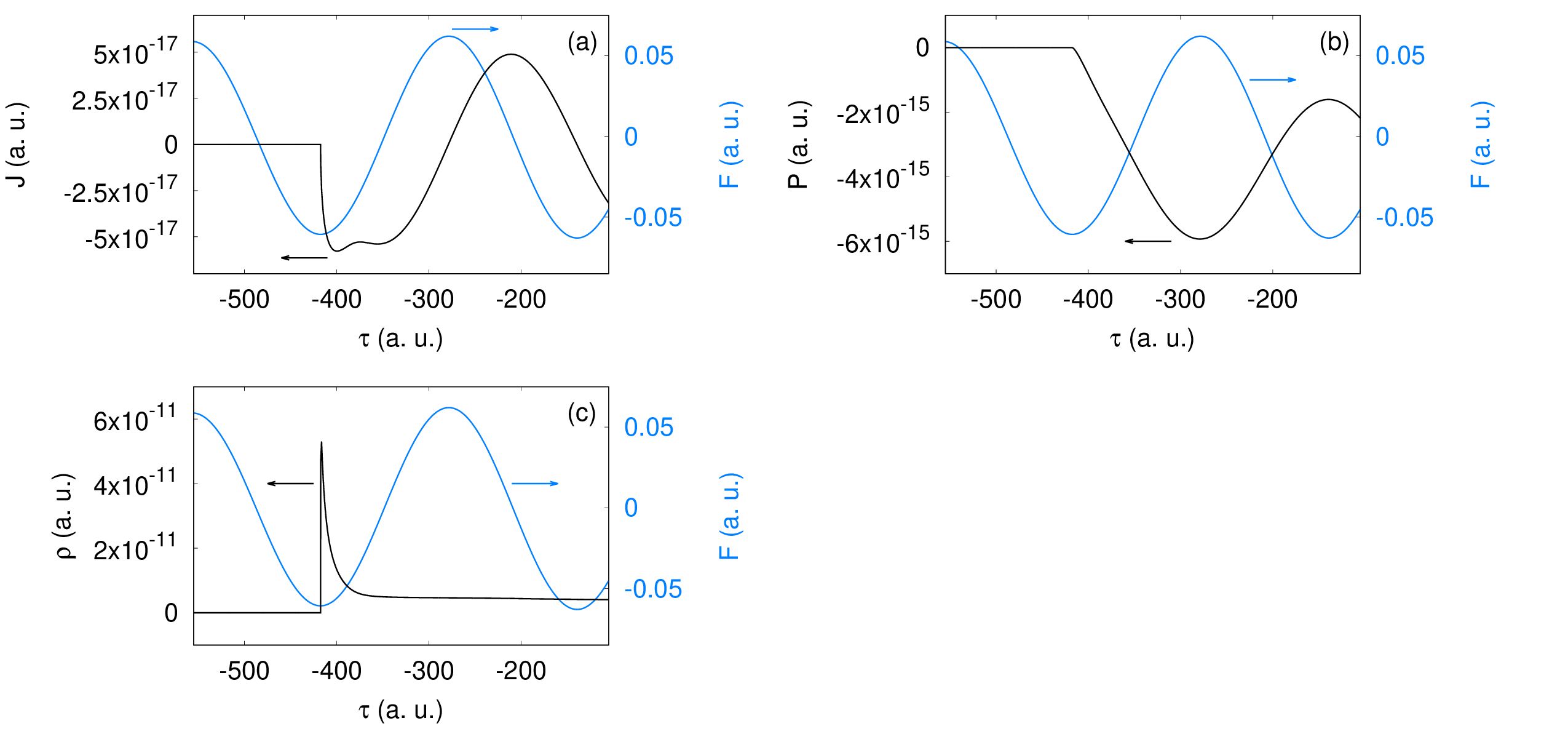}
\caption{\label{isol_1} Temporal dependencies of the plasma-induced current (a), polarization (b), and plasma density (c) for an isolated ionization event with account of the Coulomb potential. In all cases, atomic hydrogen is considered, and blue curves represent the pump electric field.}
\end{figure}

To understand the above differences in behavior, we start with the study of the time-dependent characteristics of the photoionization for a single ionization event (single saddle point $t_s$), as shown in Fig. 3 and Fig. 4 with and without the Coulomb effect, correspondingly. The physical picture illustrated in Fig. 3 is quite trivial: after the ionization event at -415 a.u., an electron moves far from the ion under the action of the electric field, as can be seen by free oscillations of  current and polarization in the electric field. The time-dependent density of the ionized states, as shown in Fig. 3(c), exhibits a single jump at the ionization time of -415 a.u.
\begin{figure}
\includegraphics[width=0.45\textwidth]{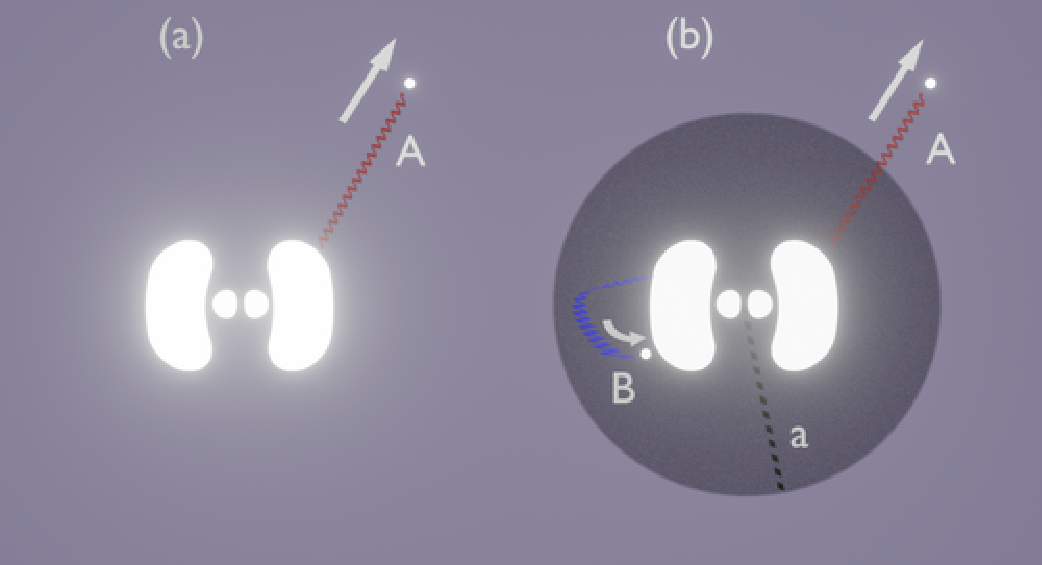}
\caption{\label{fig:epsart} Schematic representation of electron motion in short-range (a) and long-range (b) potential.}
\end{figure}

\begin{figure}
\includegraphics[width=0.5\textwidth]{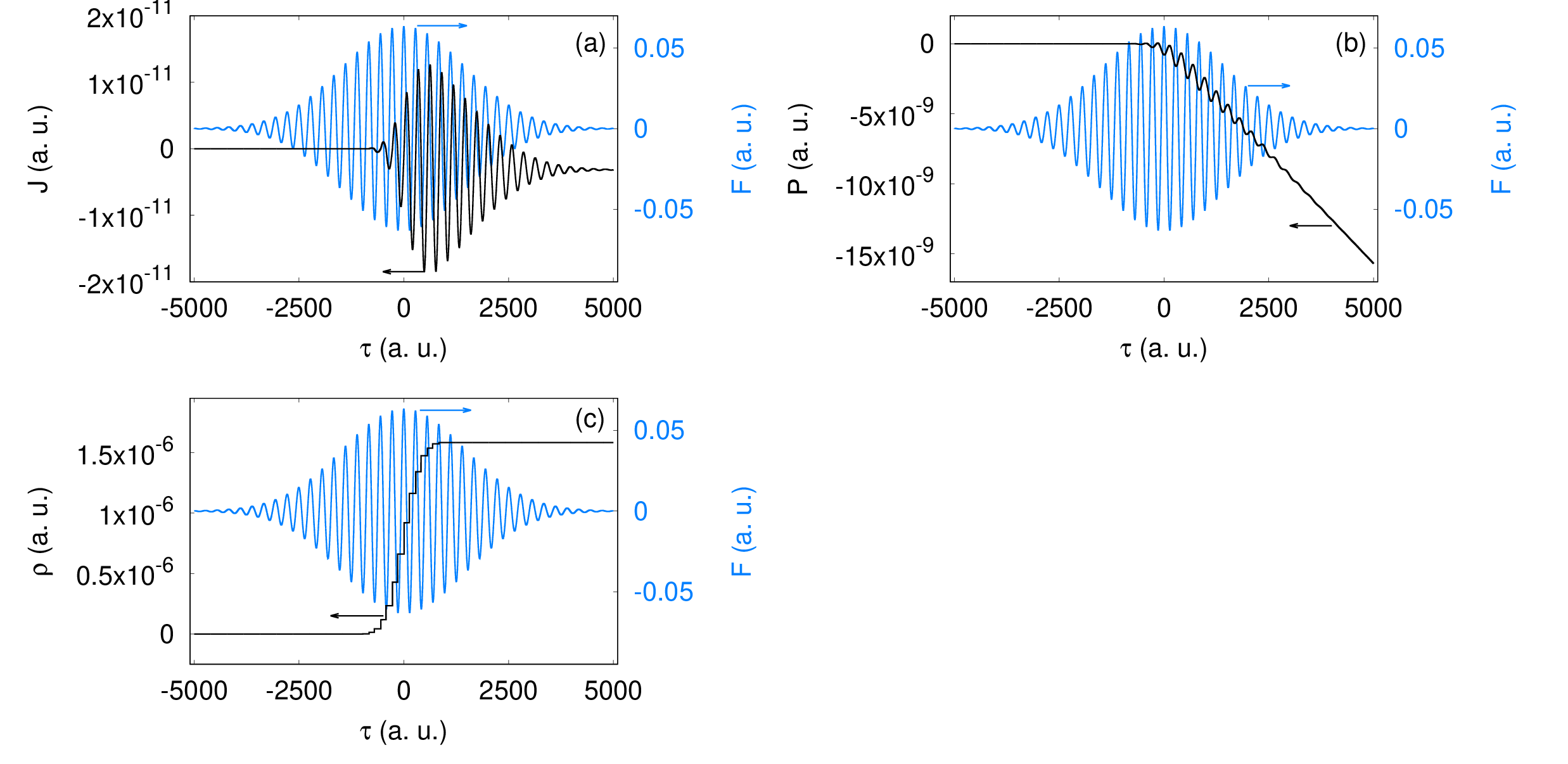}
\caption{\label{glo0} Temporal dependencies of the plasma-induced current (a), polarization (b), and plasma density (c) for a sum over all ionization events without the Coulomb potential. In all cases, atomic hydrogen is considered, and blue curves represent the pump electric field. }
\end{figure}

The physical picture is quite different when taking into account the Coulomb potential, as shown in Fig. 4. As shown in Fig. 4(a), from -415 to -380 a.u. of time electron moves roughly in phase with the electric field, which corresponds to real-valued conductivity and nonlinear loss. Electron motion during this time has no counterpart in the case of short-range potential. Only after -380 a.u. of time, the motion of the electron becomes free and it oscillates under the action of the electric field, just as happened in the case of a short-range potential shown in Fig. 3. The density of the ionized states, depicted in Fig. 4(c), exhibits a rise at -415 a.u. of time followed by a significant decrease, with stationary value reached around -380 a.u. of time. 

The physical picture of the electron motion is schematically represented in Fig. 5. Without the Coulomb potential, as shown in Fig. 5(a), the free electron motion commences immediately after it leaves the core, as shown by the trajectory A. In contrast, for the long-range Coulomb potential, after leaving the core, the electrons enters an intermediate stage (trajectory B) whereby it dwells in the area of the influence of the Coulomb potential within the radius $a$ from the core. During this intermediate phase, the electron moves almost in phase with the electric field, leading to enhanced nonlinear absorption. However, only a fraction of electrons which have entered this "nearly-free" intermediate phase will eventually become truly free; there exists a significant probability that the electron will return to the core and become completely bound again, as shown in Fig. 5(b).  This return also manifests itself in Fig. 4(c), where one can see the drop of the ionized states density between the -415 and -380 a.u. of time. Such nearly-free electrons contribute to the nonlinear loss and negative imaginary part of the susceptibility during their almost-free motion. However, they do not contribute to the real part of the susceptibility and to the plasma defocusing, since they do not remain free after the interaction. These electrons are also responsible for the "broken" proportionality between the real and imaginary part of the susceptibility, as shown in Fig. 1(b). Their presence is the main reason why it is not sufficient to describe the plasma response influenced by the  long-range Coulomb potential by a simple modified ionization rate, as it was done in Ref. \cite{Schuh}.
\begin{figure}
\includegraphics[width=0.5\textwidth]{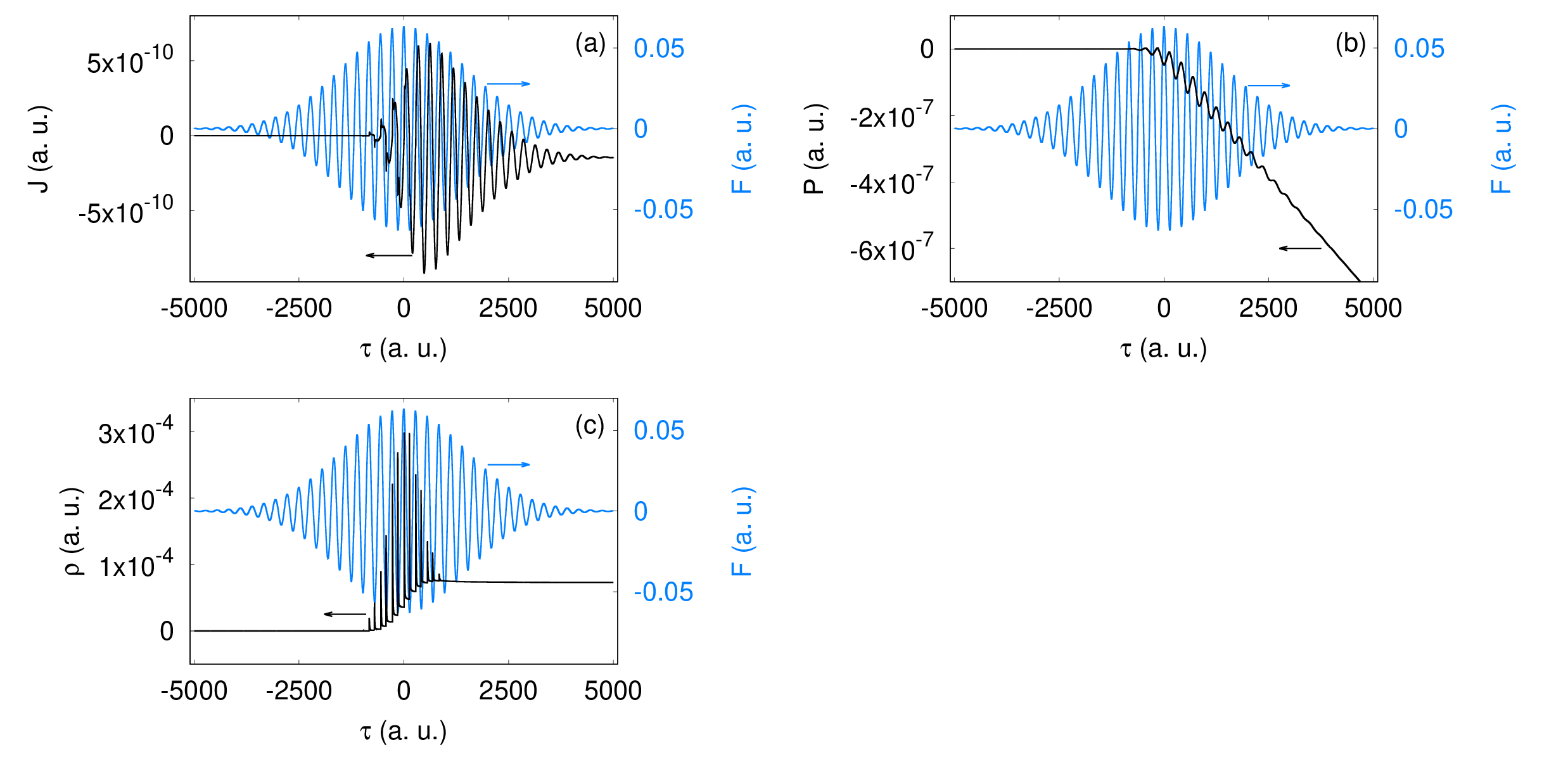}
\caption{\label{glo1} Temporal dependencies of the plasma-induced current (a), polarization (b), and plasma density (c) for a sum over all ionization events without the Coulomb potential. In all cases, atomic hydrogen is considered, and blue curves represent the pump electric field. }
\end{figure}

Now shifting our attention to the picture including all the ionization events and all the canonical momenta, we illustrate the role of the almost-free electrons by comparing the case without and with the long-range Coulomb potential in Figs. 6 and 7 respectively. The difference is most easily seen in Fig. 6(c) and Fig. 7(c), whereby a steady growth of the plasma density is visible for the case without the Coulomb potential. In contrast, for long-range Coulomb potential each increase of the plasma density is followed by a significant recombination of the almost-free electrons by the ion and corresponding drop of plasma density. The motion of almost-free electrons is also visible in Fig. 7(a), where one can see many spikes of current which are in phase with the electric field. Such spikes correspond to electrons which contribute to the nonlinear absorption but not to the plasma defocusing, i.e. the real part of the susceptibility. This provides an explanation also to the starting finding of this paper, namely the intensity-dependent ratio of the real and imaginary part of the susceptibility, as shown in Fig. 1[b].

\begin{acknowledgments}
Authors acknowledge financial support from German Research Council, projects HE2083/24-1 and HU1593/11-1.
\end{acknowledgments}

\section{Conclusion}
 In this paper, we have applied the analytic R-matrix method for the calculation of the frequency-dependent nonlinear plasma contribution to the nonlinear refractive index and the nonlinear loss. We predicted that due to the plasma contribution the real part of the susceptibility is more than one order of magnitude higher if the long-range Coulomb potential is included compared to short-range interaction. We have shown how an additional contribution to the imaginary part of the nonlinear susceptibility arises from the long-range Coulomb potential of the atomic core. We have analyzed the dynamics of the electron wavefunction during different stages of the ionization process both with account of the  Coulomb potential and without the Coulomb potential. We have demonstrated that the additional contribution is caused by nearly-free electrons in states very close to continuum and their motion in phase with the electric field.

\end{document}